\documentclass{IEEEtran}
\usepackage{graphicx}        
\usepackage{subfigure}
\usepackage{amsmath}
\usepackage{balance}
\usepackage{placeins} 
\usepackage{flafter}  
\usepackage{float}
\usepackage{moreverb}
\usepackage{url}
\usepackage{marvosym}
\usepackage{algorithmic}
\usepackage{algorithm}

\begin{document}

\title {Reliable Group Communication Protocol for Internet of Things}
\author{\IEEEauthorblockN{Mauro Conti, Pallavi Kaliyar,
Chhagan Lal}\\
\IEEEauthorblockA{Department of Mathematics, University of Padova, Italy\\
\{conti, pallavi, chhagan\}@math.unipd.it}
}

\maketitle

\begin{abstract} Routing Protocol for low power and Lossy networks (RPL) is a standardized routing protocol for low power and lossy networks (LLNs) such as Internet of Things (IoT). RPL is designed to be a simple (but efficient) and practical networking protocol to perform routing in scalable IoT networks that consists of thousands of resource constrained devices. These tiny intercommunicating devices are currently in use in a large array of IoT application services (e.g., eHealth, smart agriculture, smart grids, and home automation). However, the lack of scalability, the low communication reliability, and the vulnerability towards various security threats still remains significant challenges in the broader adoption of RPL in LLNs.
\par In this paper, we propose \textit{RECOUP}, a reliable group communication routing protocol for IoT networks. RECOUP efficiently uses a low-overhead cluster-based multicast routing technique on top of the RPL protocol. RECOUP increases the probability of message delivery to the intended destination(s), irrespective of the network size and faults (such as broken links or non-responsive nodes), and in the presence of misbehaving nodes. We show that the cluster-based routing mechanism of RECOUP remains robust in presence of various topology (i.e., rank and sybil) and data communication (i.e., blackhole, wormhole, and jamming) attacks targeting the IoT networking infrastructure. An implementation of RECOUP is realized in Contiki. Our results show the effectiveness of RECOUP over state-of-art protocols concerning packet delivery ratio to 25\%, end-to-end delay down to 100 ms, low radio transmissions required for per packet delivery to 6 mJ, and most importantly, it improves the robustness and scalability of data communication process in the whole network.

\end{abstract}


\section{Introduction}
\label{sec:intro}

In Internet of Things (IoT) networks, sensors collect data and send it to base stations or actuators for storage, processing, and service creation~\cite{Fuqaha2015}. The IoT devices are usually clubbed together, logically, into groups based on their functionalities and utility. These groups usually: (1) collect and send data to the base station/actuators, and (2) receive specific commands/requests from the base station to perform necessary actions. In particular, IoT network consist of constrained sensor devices (also called \textit{motes}) that create a Low-power Wireless Personal Area Network (LoWPAN) in which communication is done using a compressed Internet Protocol Version 6 (IPv6). LoWPAN over IPv6 (i.e., 6LoWPAN) uses the IEEE 802.15.4 as the data-link and physical layer protocol~\cite{Kushalnagar2007}~\cite{Shelby2012}. 

\par For routing in  resource constrained networks such as IoT, the Routing Protocol for Low-power and Lossy Networks (RPL)~\cite{Winter2012} is considered as an idle routing solution. RPL mainly supports point-to-point (P2P) communications (i.e., unicast), however it provides an optional support for multicast routing. The majority of real world IoT applications such as home automation and security management, environmental monitoring, and smart energy monitoring would be more benefited, if point-to-multipoint (P2M) (i.e., multicast) routing will be in use for data dissemination and for machine-to-machine (M2M) communication. Due to RPL's unicast routing nature, the low scalability, high propagation delay, and high energy consumption becomes significant routing issues~\cite{Hui2012}. Furthermore, in RPL, the availability of a single route between source and destination pairs cause major challenges concerning communication reliability and security in the network. Hence, it leaves the door open to multiple security threats wherein an adversary can disrupt the routing process by just compromising a single node in the network~\cite{rplsurvey2017}. Therefore, our work aims to improve communication in IoT networks, mainly in terms of routing reliability (i.e., low delay and high network throughput), scalability, and security.

\subsection{Motivation and Contributions}

Due to the shared wireless channel and lack of any physical protection and tamper resistance (i.e., nodes can be easily captured, tampered, or destroyed by an attacker), LLNs are easily threatened by an array of security attacks. These attacks primarily disrupt network protocols and interfere with the data communication process. For instance, an attacker can exploit vulnerabilities in RPL's functionality to launch specification-based attacks such as Rank attacks~\cite{Perreyy2016}, DAO/DIA attacks~\cite{Tsao2015AST}, and Version number attack~\cite{Anth2014} Furthermore, the lack of support for mobility in RPL makes it vulnerable to mobility-based attacks such as sybil attack as well as it increases packet loss rate due to broken and short-lifetime links.

\par In this paper, we propose a reliable (i.e., able to cope with link and node failures) and robust (i.e., able to cope with security attacks) group communication protocol (RECOUP) for efficient data communication in LLNs such as IoT. The key functionalities of RECOUP are as follow: (i) virtual clusters creation on top of RPL's logical Destination Oriented Directed Acyclic Graph (DODAG) topology, (ii) perform upward and backward multicast routing in DODAG by using RPL's storing mode of operation (also called as MOP3), and (iii) optimized inter-cluster routing for quick dissemination and reliable delivery of multicast packets. These functionalities of RECOUP lead to the low data packet propagation delay, high packet delivery ratio, and minimal effect of various topology and communication attacks in the network. This paper is an extension of our previous work called \textit{REMI}. The basic idea behind REMI along with the initial simulation results were first presented in~\cite{conti2017}. We have extended REMI along several important dimensions which includes the following: 
\begin{itemize}
\item the functionality of REMI's routing mechanism has been extended to optimize the network overhead and energy consumption. In particular, we have added new packet forwarding technique at root node, and the inter-cluster routing has been optimized to control the duplicates and broadcast messages, 
\item the working methodology of the RECOUP is updated according to the extensions done on various modules of REMI, and an example scenario has been added to better explain the RECOUP functionalities and benefits, and 
\item the evaluation section is significantly enhanced by including additional results obtained on large number of target scenarios with varying network size, network load, and number of attacker nodes. Additionally, the results analysis are extended to evaluate the proposed protocol for various new network metrics that were absent in REMI. 
\end{itemize}


\par In summary, the main contributions of this work are as follow.
\begin{itemize}
\item We design and fully implement \textit{RECOUP}, a novel and reliable multicast routing protocol for efficient and robust data communication in IoT networks. RECOUP makes efficient use of RPL control messages to implement its optimized cluster based horizontal routing mechanism, thus it avoids additional memory and control overhead in its execution process. Our detailed discussion on RECOUP's data communication reliability and robustness in resisting an array of security threats in different IoT networking scenarios show the major advantages of RECOUP. The paper also report the key implementation issues of RECOUP which includes energy consumption and memory requirements on network nodes while running the RECOUP protocol. 

\item We perform a comprehensive performance evaluation of RECOUP concerning various network metrics such as end-to-end delay, packet delivery ratio, average path cost, energy consumption, and memory requirements. To shows the efficiency of RECOUP regarding communication reliability and robustness to security threats, the result evaluation is done with varying network size and in the presence of attacker nodes in the target scenario. Furthermore, to show its effectiveness, we compare RECOUP with the following RPL based state-of-the-art multicast routing protocols: (i) ESMRF~\cite{Fadeel2015}, a enhanced stateless multicast RPL-based forwarding protocol, and (ii) BMRF~\cite{Lorente2017}, a bidirectional multicast RPL forwarding protocol. The implementation is done in \textit{Cooja}, the Contiki network emulator~\cite{Romdhani2016}, which is widely used for deploying energy-constrained and memory-efficient LLNs. We make available\footnote{https://github.com/pallavikaliyar/RECOUP} an open-source implementation of RECOUP along with all the source code to the research community.
\end{itemize}

\subsection{Organization}
The rest of this paper is organized as follow. In Section~\ref{sec:related_works}, we discuss the state-of-the-art IoT routing protocols and techniques that addresses the security and reliable data communication issues in IoT networks. In Section~\ref{sec:SeMI}, we present the system and adversary model, and the design and implementation details of RECOUP along with its working methodologies. In Section~\ref{sec:result}, we present the detailed performance evaluation of RECOUP in terms of various essential metrics using the \textit{Contiki Cooja} emulator. Finally, Section~\ref{sec:conclusions} concludes the work done in the paper along with the directions of future work.

\section{Background and Related Work}
\label{sec:related_works}
\noindent

In this section, first we present a brief overview of the state-of-the-art IoT routing protocols that are designed for 6LoWPAN based wireless sensor networks. Then we discuss the related work concerning security and reliable data communication in RPL-based IoT networks.   

\subsection{Routing Protocol for Low Power and Lossy Networks (RPL)}
\label{sec:RPL}

\par RPL~\cite{Winter2012} creates a virtual routing topology called Destination-Oriented Directed Acyclic Graph (DODAG) on top of the underlying random physical topology. DODAG is a directed graph with no loops, oriented towards a root node (e.g., a LLN/border router). Each node receives a rank ID whose value depends on its distance from the root. In DODAG, each node by default have multiple parents towards the root, however, only a preferred one which is selected based on routing metric and objective function is used for forwarding data packets, while the others are kept as backup routes. The structure of DODAG naturally supports multipoint-to-point communication in RPL, which provides communication from the nodes to the root with minimal routing state. The DODAG topology is created and maintained via ICMPv6 control packets know as DODAG Information Objects (DIO). Each node is RPL advertises DIO messages, which contains the link and node metrics (e.g., expected transmission count (ETX), residual energy) and an objective function that are used by each node to select the preferred parent among the candidate neighbors. To maintain the DODAG, DIO packets are rebroadcast by each node based on the Trickle algorithm~\cite{Levis2011}, which is an adaptive technique that tries to achieve a balance between reactivity to topology changes (fast convergence/recovery) and control overhead (energy consumption). In particular, Trickle ensures that DIOs packets are rebroadcast at slow pace when the network is stable, and aggressively when it is unstable. DIO packets are also transmitted upon request when a node receives a DODAG information solicitation (DIS) packet, which could be sent by a new node that wants to join the DODAG. 

\par Apart from multipoint-to-point communication, the RPL supports point-to-multipoint and point-to-point communications in two modes called storing and non-storing modes. In storing mode (table-driven routing), the non-root nodes store the routing information about all the descendant nodes in its subtree (i.e., sub-DODAG), while in non-storing mode (source routing) the routing information about all the nodes is stored at the root. In both the modes, the routing information is collected using Destination Advertisement Object (DAO) control packets, which are transmitted by each node in the network to announce itself as a possible destination to the root. DOA packets are propagated towards the root, via a parent, therefore establishing ``downwards’’ routes along the way. The detailed working of RPL and its features are out of the scope of this paper. Therefore, we direct the interested readers to more comprehensive literature given in~\cite{Winter2012} and~\cite{Kim2017}.

\par The first extension that uses the RPL functionality is proposed in~\cite{Oikonomou2013}, and it is called Stateless Multicast RPL Forwarding (SMRF). In SMRF, nodes only process the multicast packets which are coming from their preferred parents, hence SMRF only allows the forwarding of multicast packets in downward direction in the RPL DODAG tree. The extensions of SMRF called Enhanced Stateless Multicast RPL Forwarding (ESMRF)~\cite{Fadeel2015} in which sources of multicast traffic encapsulates their multicast packet in an ICMPv6 delegation packet and send it to the root for forwarding, and Bi-Directional Multicast Forwarding Algorithm (BMFA)~\cite{Papadopoulos2017}, which improves SMRF and enable multicasting in upward and downward directions. Finally, authors in~\cite{Lorente2017} propose the \textit{Bidirectional Multicast RPL Forwarding} (BMRF) protocol, which fully utilizes the potential of RPL's non-storing mode to overcome various disadvantages of SMRF. In BMRF, when a node wants to send a multicast message, it performs the bidirectional forwarding. BMRF provides a choice for Link Layer unicast, broadcast, or mixed mode to forward a multicast packet at a parent node. Link Layer unicast or broadcast depends upon the number of interested children and mix mode depends upon whether the number of interested children are larger than a pre-defined threshold value. In addition, BMRF added one more new feature that allows a node to un-subscribe itself from a multicast group by sending a DAO message to the preferred parent. The main advantages of BMRF includes that it avoids duplicates and overheads, there is no delivery disorder, and it enables multi-sourcing, i.e., at a single time in a network more than one source node can send multicast messages to the same multicast destination address. However, the BMRF also possess a set of disadvantages such as higher energy consumption, latency, and lower communication reliability and security.

\subsection{Security threats to RPL and its related protocols}

As RPL or an extension of RPL are the most used routing protocols in IoT networks. We now briefly discuss the security challenges that these protocols might face during the routing process. Authors in~\cite{Weekly2014} propose a sinkhole attack mitigation method that integrates rank authentication with parent fail-over. The proposal uses DIO message along with the one way hash function technique for rank authentication. The root node generates hash value by selecting random numbers, and broadcast these values through DIO messages. When the root node again broadcast the initially selected random number securely then intermediate nodes can verify its parent rank using the intermediate hops number. In~\cite{Khan2013} authors propose a Merkle tree authentication based solution which can be used to prevent wormhole attack on RPL protocol. In this proposal, the RPL tree is formed in the reverse direction by using the node ID and public key which are used to calculate the hash values. After the Merkle tree formation, the authentication for any node starts from the root node and if any intermediate node fails to authenticate, a possible wormhole is detected.

\par Authors in~\cite{misbehave2018} investigates the forwarding misbehaviour (i.e., selective packet discarding) and propose a countermeasure of the same in LLNs running with RPL protocol. The basic idea is to monitor the forwarding (mis)behaviour of each node to observe the packet loss rate, and then compare the packet loss rate of the parent node with the neighbor nodes. To ensure that the packet loss is due to misbehaviour and not due to bad channel quality, the nodes use one time retransmission techniques. Similarly, using the monitoring information of the nodes about the data packets forwarding, a trust-based intrusion detection system based on RPL is presented in~\cite{sybil2018}, to countermeasure mobile sybil attacks in IoT networks.  

\par In ~\cite{Ahmeds2016}, authors propose solution to mitigate black hole attack in RPL, identifying the malicious node by using a mechanism similar to \textit{watchdog}, in which the neighbour nodes keep record of a nodes activities and analyze it find any malicious behaviour. Authors in~\cite{Dvir2011}~\cite{Perreyy2016} addresses the \textit{rank}\footnote{An attacker decreases its rank to spoil the routing topology and attract traffic from neighbor nodes, which degrades packet delivery performance when combined with blackhole, wormhole, or selective forwarding attacks.} attack, which is an attack specific to RPL. VeRA~\cite{Dvir2011} effectively fixes the vulnerabilities caused by the false rank of a node and the DODAG version number dissemination. VeRA does it by adding reverse hash chaining to DIO messages due to which receivers shall be enabled to verify the advertised hierarchy. However, in~\cite{Perreyy2016} authors show that VeRA remains vulnerable to rank attacks by forgery and replay, and they propose TRAIL (Trust Anchor Interconnection Loop), which aims to discover and isolate bogus nodes. The key idea is to validate upward paths to the root using a round trip message. This is achieved without relying on encryption chains (as in VeRA), in TRAIL a node can conclude rank integrity from a recursively intact upward path. Recently, authors in~\cite{pals2018} propose a secure and scalable RPL routing protocol (SPLIT) for IoT networks. SPLIT uses a lightweight remote attestation technique to ensure software integrity of network nodes, thus ensures their correct behaviour. To avoid additional overhead caused by attestation messages, SPLIT piggybacks attestation process on the RPL's control messages.   

\section{Our Proposal: RECOUP}
\label{sec:SeMI}

In this section, first we present the details of the system and adversary models on which RECOUP is implemented and evaluated. Then we discuss the working methodology of our proposed protocol, i.e., RECOUP, along with its design considerations, characteristics, and routing process. 

\subsection{System Model}
\label{sys_model}

In our work, we assume that the system model has the following properties.

\begin{itemize}

\item The target network consists of a set $D = \{D_1, D_2, ... D_n\}$ of size $n$ resource constraint IoT nodes (i.e., sensors and actuators). These nodes are static within the IoT network area. We consider that all the nodes are homogeneous in terms of resources, but could be different in terms of their functionalities depending upon the configured sensor type such as temperature, illumination, audio, pressure, to name a few. Nodes with similar functionalities are grouped together to form a multicast group in the network. All the nodes are configured using the standard layered protocol stack of IoT. At network layer the nodes use RPL MOP3 (i.e., storing with multicast support) over IPv6 as a routing protocol for data communication.         

\item At start, the $n$ nodes are deployed in a random fashion, and the RPL creates a virtual DODAG on top of the physical network topology. Apart from the $n$ nodes, the network also has a resourceful nodes called LLN border router (LBR) which acts as the root for the DODAG(s) in the network. A network could have more than one DODAG represented by different DODAG IDs ($DID_i$) and different root nodes. Each node in the DODAG has a $rank$ value which specify its level in DODAG, i.e., distance from the root. The rank of the root is set to $0$, and the rank associated with a node increases with its distance from root.

\item In RECOUP, each DODAG is divided into a set of clusters, and the nodes having rank $1$ will act as the clusterheads. For instance, the nodes with IDs 1, 5, 7, 12, 14, 15, and 16 in Figure~\ref{Fig:RECOUP} will act as clusterheads within that particular DODAG. Each cluster is represented by a unique ID ($CID_{i}$). It can be seen from the Figure~\ref{Fig:RECOUP} that nodes in DODAG are arranged in a parent and child structure, each parent store information about its children which includes their subscription for a multicast group among other data. A node could be subscribed for more than one multicast group depending upon the usage requirements of IoT application running on top of the network.   

\item Multicast routing is used to send data messages to a group of nodes with similar functionalities. However, the network also supports point-to-point and multipoint-to-point routing. The source of the multicast message could be the root node or a member of a multicast group. Data exchanged between two communicating nodes that are not within each others radio range will be forwarded by intermediate nodes. All nodes are capable of operating in upward, downward, and inter-cluster routing modes and sensing/actuating modes.

\item RECOUP uses the following additional or enhanced data structures at different nodes (i.e., root, clusterhead, and cluster members) in the DODAG. 

\begin{itemize}
\item \textit{Neighbour Table ($N_{tab}$):} The $N_{tab}$ can be simply created by extending the functionalities of the traditional RPL routing tables which a node stores when routing is done in RPL's storing mode. Usually, the routing table at a node stores the information for all its descendant nodes. Each $N_{tab}$ entry stores the following information about a neighbour node (say $N$): (i) cluster ID ($CID_{i}$) of $N's$ cluster, (ii) node ID of $N$ ($NID$), and (iii) rank of $N$. As stated before that we implement $N_{tab}$ on top of the existing \textit{information/routing table} that already exists at all the nodes in the network. The additional information that RECOUP adds is the $CID$ of a node and the entires for the neighbour nodes which are not the descendants, thus keeping the low memory consumption. The $N_{tab}$ is associated with a timer called $TrickleTimer$ ($TT$), and the nodes update the $N_{tab}$ with new network information once this timer expires. The value of the timer is set by the network administrator depending upon the RPL DODAG reformation i.e., if any new node joins the existing DODAG or any existing node changes its parent node within the DODAG.

\item \textit{Duplicate Detection Table ($DD_{tab}$):} RECOUP uses $DD_{tab}$ at root node for two purposes: (i) to check for the duplicates because in RECOUP the same packet is travelling towards root through multiple clusters, and (ii) to hold the received packet for a variable time duration while waiting for all the duplicates to receive from multiple clusters. The $DD_{tab}$ consists of a set of entries, where each entry has the following information about the received multicast packet ($MP_i$): (i) source address ($S_{id}$), (ii) destination address ($D_{id}$), (iii) set of cluster ID(s) from which $MP_i$ is received ($CID_{set}$) so far i.e., ID of the cluster from which the packet or its duplicate has been received, (iv) packet sequence number ($N_{seq}$), (v) forwarding timer ($F_{time}$), and (vi) a buffer to hold the $MP_i$ until the associated $F_{time}$ expires. Upon expiration of the $F_{time}$, the packet is processed and the entry is removed from the $DD_{tab}$. The tuple $<$ $S_{id}$, $D_{id}$, $N_{seq}$ $>$ is used to discard duplicates. We have implemented $DD_{tab}$ as a dynamic link list at LBRs or root only. Additionally, same as the traditional RPL, each node in RECOUP also stores the $DD_{tab}$ whose function is limited to just detect and discard the duplicate packets, i.e., the entries in $DD_{tab}$ at non-root nodes only consists of the tuple $<$ $S_{id}$, $D_{id}$, $N_{seq}$ $>$.
\end{itemize}
\end{itemize}

\subsection{Adversary Model}
The use of IoT networks in a large array of user-centric applications make these networks a high profit target for the adversaries. Hence, the adversaries would try their best to equip themselves with advanced equipment, which means they would have few technical advantages over the IoT nodes. In our target IoT network, an attacker is interested in minimizing the connectivity of the network to prevent the LBR or members of a multicast group from detecting important events, thus impairing their decision making system. To achieve this goal, the attacker selects at each time $t$ a node to compromise from the set $D$. In fact, the attacker chooses the node which maximizes the adverse impact on the IoT services running on top of the networking infrastructure. For instance, the attacker could launch a rank attack followed by a blackhole attack to create a large network partition.

\par In our target IoT network, the adversaries are assumed to have the following characteristics: 
\begin{itemize}
\item The adversary is resourceful, and it could perform the rank, jamming, blackhole, eavesdropping, and wormhole attacks. To launch the aforementioned attacks, it can compromise an existing node or it can be part of an existing network as a new node. However, we assume that the adversary cannot compromise the LBR (i.e., DODAG root).
\item The adversary will not interfere with the proper functioning of the network such as modifying the data packets, generating new messages, destroying network devices, and tempering with the key distribution and management operations. It is because such activities can be easily detected by an IDS and could put the adversary at risk of being caught~\cite{IDS2017}.
\end{itemize}

\begin{table}[h!]
\centering
\caption {Symbol table}
\scalebox{0.87}{
\begin{tabular}{|p{3cm}|p{5cm}|} \hline
\textbf{Symbol} & \textbf{Meaning} \\ \hline
$X$ & node in DODAG  \\ \hline
$P$ & parent of a node \\ \hline
$TX_{MP}$ & link layer transmission time of a multicast packet \\ \hline
$X_r$ & rank of node X  \\ \hline
$MP_i$ & $i^{th}$ multicast data packet \\ \hline
Src~($MP_i$) & source of $MP_i$ \\ \hline
$MG_i$ & $i^{th}$ multicast group in network  \\ \hline
$X_{CID}$ & ID of X's cluster  \\ \hline
$IC$ & set of interested children of a node \\ \hline
$DD_{tab}$ & duplicate detection table \\ \hline
$F_{time}$ & $MP_i$ hold time at LBR\\ \hline
$CID_{visit}$ & set of cluster IDs at LBR from which $MP_i$ is received \\ \hline
$C2C$ & inter cluster routing set/reset flag\\ \hline
X~($N_{tab}$) & neighbour table at X  \\ \hline
$Pkt_{drop}$ & maximum hop-count for $MP_i$ in inter-cluster forwarding \\ \hline
$NC_{ID}$ & ID of $i^{th}$ neighbour cluster \\ \hline
$N_j$ & $j^{th}$ neighbour from a $NC$  \\ \hline
$C_{travel}$ & set of clusters travelled by $MP_i$ via inter-cluster forwarding\\ \hline  
\end{tabular}}
\label{tab:symbol} 
\end{table}

\subsection{RECOUP Design Considerations}
\label{sec:RECOUP_Cons}

Below is the list of design considerations along with their functioning details that were taken into account while designing RECOUP protocol. 

\begin{itemize}
\item \textit{Cluster Formation:} In RECOUP clusters are formed along with the creation of the DODAG tree, and it extends the RPL MOP3 protocol to keep the basic functionality of DODAG creation intact. In particular, clusters are created virtually on top of the DODAG tree. Each cluster within the DODAG could be seen as a separate DODAG with clusterhead acting as its root. We limit the rank for clusterheads to 1, i.e., only the children of the root in the DODAG can act as clusterheads (please refer to Figure~\ref{Fig:RECOUP}). Hence the number of clusters in a network will be equal to the number of children of a DODAG root. The root will assign a unique ID to each cluster called cluster ID (CID), and all the nodes that belongs to the same cluster will share a common CID. 

\item \textit{Information Storage:} As the nodes in RECOUP will be configured with RPL MOP3 mode, each node will store the essential information needed to route the messages in upward and downward routes in their $N_{tab}$. Additionally, the nodes will also store the duplicate detection table $DD_{tab}$ as described in Section~\ref{sys_model}. A node could easily learn about its current neighbour set by using the information that it receives through one-hop periodic DIO messages and it will store this additional information in its $N_{tab}$. These messages are sent by the nodes to communicate their existence in network and to maintain the DODAG topology. We add an additional field in the DIO header to carry a node's cluster ID which is required to extend the functionalities of $N_{tab}$.

\item \textit{Duplicate Avoidance:} One way to avoid the duplicates is that each node check the received packet for possible duplicate before processing it. However, checking every packet will increase the energy consumption and end-to-end delay in the network. In RECOUP, to minimize the duplicate messages the following optimization's are included: (i) the inter-cluster forwarding is limited by using a threshold hop-count value, (ii) a node will send a message to only one neighbour from the group of neighbours if this group of neighbour belongs to the same cluster, (iii) we use low transmission range (i.e., 25 meters) for data communication to reduce both, the overlapping in neighbouring clusters and the re-transmissions required to forward a message to its next hop, and (iv) a node will not forward a data packet in the direction from which it has been received, i.e., to the children or parent or cluster. Additionally, in RECOUP, the data packets are always sent as a unicast. The details about the aforementioned duplicate avoidance steps will be presented in Section~\ref{sec:RECOUP_routing}. 

\item \textit{Multi-directional Forwarding:} The multicast packet will be is forwarded in all directions, which includes upwards (i.e., preferred parent), downwards (i.e., interested children) and neighbour (i.e., neighbour that is a member of different cluster) nodes. This feature of RECOUP plays an important role, specifically to decrease the propagation delay and to increase the network scalability, reliability and security. 

\end{itemize}

\subsection{RECOUP: Reliability in Data Communication and Resistance against Security Threats}
\label{sec:RECOUP_routing}

In this section, we present the working methodology of \textit{RECOUP} protocol, which is mainly divided into two phases. The first phase consists of the DODAG and cluster formation. This phase also includes the possible updates in the DODAG and clusters that are caused by the change in network topology triggered by node join, node leave/revocation, and node changing parent within the DODAG. The second phase consists of the cluster-based data packet routing technique. It includes packet forwarding within the cluster (i.e., intra-cluster) and in-between the neighbour clusters (i.e., inter-cluster). In this phase, a root or non-root node could send multicast messages in the network. The symbols used for explaining the routing mechanism of RECOUP are shown in Table~\ref{tab:symbol}.

\subsubsection{DODAG and Cluster Creation in RECOUP}

To minimize the convergence time of the network and to avoid additional control messages, RECOUP uses an optimized cluster formation approach. The cluster formation approach is straightforward and it fully depends on the RPL's DODAG creation. In particular, the cluster formation is done in parallel with the DODAG creation by using the following steps. 

\begin{itemize}

\item The LBR broadcasts the RPL's DAG information object (DIO) message with the required information such as DODAG ID, rank (i.e., $0$), trickle timers etc.    

\item When a node receives a DIO message from neighbours, it selects a preferred parent according to its objective function (OF). For instance, when a node receives a DIO message from LBR, it selects the LBR as its parent, and it calculate its own rank (i.e., $R_i = R_p + 1$, where $R_p$ is the parent rank). Once a node select its preferred parent, then it notifies the parent by sending a DAO message, and the parent confirm it by replying with a DAO-ACK message. In particular, while RPL uses DIO and DIS messages to create the upward routes (toward the LBR), it uses DAO messages to maintain and find the downward routes in the DODAG (from the LBR or parent node toward children or leaf node).

\item In RECOUP, a node with rank $1$ will act as clusterhead, and any node that joins a cluster (i.e., descendent of clusterhead) will use the same cluster ID that is assigned to the clusterhead. Initially, the clusterheads receive their cluster IDs from the LBR in the DAO-ACK messages.

\item Once the clusterheads have their unique cluster IDs assigned by the LBR, the clusterheads broadcast DIO messages with their own cluster ID. The nodes that receive these DIO messages will select a preferred parent, keep the received cluster ID, calculate their own rank by increasing the parent rank by $1$, and then broadcast the DIO again in the network. This process is repeated until the network constructs the routing topology (i.e., DODAG). If a new node joins the network, it could discover the nearest DODAG by sending the DIS message. When a cluster member receives a DIS message, it replies with a DIO message.

\end{itemize}

\par By executing the aforementioned steps, RECOUP creates the required clusters in parallel with the formation of the DODAG, thus minimizes both, the control overhead messages and the network convergence time. As the cluster formation is closely coupled with the RPL's DODAG creation, there is no need to perform the cluster maintenance as it happens automatically with DODAG's re-creation process. The DODAG re-creation or update is triggered either due to node changes parent or due to leaving or joining of IoT nodes in the network. 

 \begin{algorithm}[h!]
 \caption{LBR multicast packet routing process in RECOUP}
 \label{algo1}
\textbf{INPUT at a Node}: Data packet\\
	\textbf{OUTPUT}: Forward the data packet towards its destination \\
  \begin{algorithmic}[1]
	\IF {$X$ has a $MP_i$ to send to $MG_i$}
	    \IF {$X$ $\in$ $MG_i$}
	    \STATE deliver $MP_i$ up to the network stack \\
	    \ENDIF \\
	    \IF {Src~($MP_i$) $=$ $LBR$}
	    \STATE perform only downward routing
	    \STATE $LBR$ use source-routing to route $MP_i$ \\
	    \ELSE
	        \IF{$X$ $=$ $LBR$}
	        \STATE create new entry in $DD_{tab}$ for $MP_i$
	        \STATE $F_{time} \leftarrow (TX_{MP} \times X_r + \alpha)$ and associate $F_{time}$ with the entry
	            \WHILE{$F_{time}$ $\neq$ $0$}
	            \STATE update $CID_{visit}$ for each duplicate $MP_i$ 
                \ENDWHILE \\
                \IF{($IC \leftarrow IC$ $\cap$ $CID_{visit}$) $=$ NULL}
                \STATE drop the packet
                \ELSE 
                \STATE set $C2C \leftarrow 1$
                \STATE transmit $MP_i$ to members of $IC$ 
                \STATE GOTO Algorithm 2
                \ENDIF \\
            \ELSE
            \STATE GOTO Algorithm 2
	        \ENDIF \\
	    \ENDIF \\
	\ENDIF \\
\end{algorithmic}
\end{algorithm}

\subsubsection{Data Routing in RECOUP}

Due to the use of multicast routing in large array of practical IoT applications, we evaluate and analyze the performance of RECOUP mainly for multicast communications. However, RECOUP also supports unicast routing. When a source node wants to send a multicast packet using RECOUP, it transmit the packet in following three directions: (i) upward, i.e., towards LBR through its preferred parent; (ii) downward, i.e., towards interested children who are registered for the multicast group that is specified as the destination address in the transmitting packet header; and (iii) inter-cluster, i.e., toward neighbour(s) with different cluster ID. In case where the source node has multiple neighbours that belongs to the same cluster, the packet is sent to only one of the neighbour from that cluster. It is because if a single node in the cluster receives the packet, later it will be disseminated in the whole cluster. Next, we discuss the functionality of the routing mechanism of RECOUP for all possible data communication scenarios in an IoT network.

\subsubsection{Routing at LBR Node} 

Algorithm~\ref{algo1} shows the routing procedure at LBR/root node when it has a multicast data packet (say $MP_i$) to send (i.e., LBR is the source node) or to forward (i.e., LBR act as an intermediate node). If LBR is the source of $MP_i$, then it can perform the downward multicast routing by simply performing the source routing which uses the global network information stored in its routing table. In particular, the LBR send $MP_i$ to its interested children (i.e., the children that are subscribed to the destination multicast address specified in the packet), which also do the same and this process continues until the packet reaches to all the subscribed nodes of the multicast group. On the other hand, if the LBR is not the source of $MP_i$, this indicates that the packet is received from one or more of the underlying clusters. As the same packet might be traveling towards LBR from multiple clusters due to our inter-cluster routing, the LBR will possibly receive duplicate copies of $MP_i$. When the LBR receive the first copy of $MP_i$, it creates a new entry in its $DD_{tab}$. The entry contains a buffer to store the received packet along with other information as described in Section~\ref{sys_model}. Additionally, the LBR associates a timer called $F_{time}$ with each new entry. The value of $F_{time}$ is calculated by multiplying the rank of the source of $MP_i$ to the time taken to transmit a packet from one hop to the next hop. A random time value (say $\alpha$) is also added to $F_{time}$ to ensure that the LBR will receive all the duplicates of $MP_i$ from the cluster. The lower value of $\alpha$ will increase the number of duplicates in the network because the LBR might falsely forward the $MP_i$ in the clusters which already have the $MP_i$ through inter-cluster routing. Alternatively, the large value of $\alpha$ will increase the waiting time of $MP_i$ at LBR which will increase the routing end-to-end delay in network. In RECOUP, the initial value of $\alpha$ is set to $0$, and it is gradually increased in proportional to the number of duplicates received for an $MP_i$ after it is forwarded by the LBR. Specifically, for a data session, apart from the initial value of $\alpha$ which is set to 0, the subsequent values of $\alpha$ is estimated as follows.

\begin{equation}
\alpha = \alpha_{prev} + (T_{X_{MP}^{last}} - F_{time}^{prev})    
\end{equation}
Where $\alpha_{prev}$ is the previous value of $\alpha$, $T_{X_{MP}^{last}}$ is the total time by which all the copies of $MP_i$ has been received at LBR, and $F_{time}^{prev}$ is the previous hold time at LBR for $MP_i$.

\par Once the $F_{time}$ associated with an entry in $DD_{tab}$ is expired, the LBR forward the buffered $MP_i$ to the interested children (IC). The LBR will only forward the $MP_i$ towards the clusters from which it has not received the $MP_i$ because the interested members in remaining clusters have already received the $MP_i$ during inter-cluster routing. For this purpose, before forwarding the $MP_i$ to its IC, the LBR re-calculate its IC set (refer line 15 in Algorithm~\ref{algo1}. It removes the children that belongs to the clusters which have already seen the $MP_i$ in its way up towards the LBR. After re-calculation of IC, if the new IC set is empty, the LBR drops the packet. Additionally, to ensure that the inter-cluster routing will not happen in case where the $MP_i$ is travelling from the LBR to the clusters, we use $C2C$ flag bit in IPv6 header of the $MP_i$. When an intermediate node founds that the $C2C$ flag bit is set to $1$, it will perform only the downward routing as the upward routing and inter-cluster routing has already been taken place in past for this packet. Please note that in RECOUP all the hop-to-hop data packet transmissions in downward routing are done using an \textit{Optimized Forwarding Mechanism} (OFM) scheme as presented for BMRF protocol in~\cite{Lorente2017}.          

\begin{algorithm}[h!]
 \caption{Non-root node(s) multicast packet routing process in RECOUP}
 \label{algo2}
\textbf{INPUT at a Node}: Data packet\\
	\textbf{OUTPUT}: Forward the data packet towards its destination \\
  \begin{algorithmic}[1]
  \IF {$X$ has a $MP_i$ to send to $MG_i$}
	\IF {$X$ $=$ Src~$MP_i$}
	    \STATE $Pkt_{drop} \leftarrow X_r$  \\
	    \STATE $C_{travel} \leftarrow X_{CID} \cup NC_{ID}$  \\
	    \STATE X forward $MP_i$ towards $X_P$, $X_{IC}$, and $X(N_j)$ here $1$ to $j$ neighbors of $X$ each with different $CID$ \\
	\ELSE
	\IF{$X$ receives $MP_i$ from $X_P$}
	    \STATE perform only downward routing
	    \IF{C2C $=$ $0$}
	    \STATE forward $MP_i$ to $P$, $IC$, and $N_j$
	    \ELSE 
        \STATE forward $MP_i$ to $IC$\\	
	\ENDIF \\
	\ENDIF \\
	\ENDIF \\
    \IF {$X_P$ $\lor$ $X_{IC}$ $\lor$ $X(N_j)$ receives $MP_i$}
                \IF{routing $=$ inter-cluster $\lor$ routing $=$ within $X's$ cluster}
                \WHILE{$Pkt_{drop}$ $\neq$ $0$}
                \STATE $Pkt_{drop} \leftarrow Pkt_{drop} - 1$
                \STATE Update $C_{travel}$ by adding $CIDs$ of $N_j$ that are not present in $C_{travel}$
                \STATE forward $MP_i$ to $P$, $IC$, and $N_j$
                \ENDWHILE
                \ELSE
                \STATE forward $MP_i$ to $P$ and $IC$
                \ENDIF \\
                \IF{LBR receives $MP_i$}
                \STATE GOTO Algorithm 1\\	
                \ENDIF \\
	\ENDIF \\
  \ENDIF \\
\end{algorithmic}
\end{algorithm}

\subsubsection{Routing at non-LBR Node}

Algorithm~\ref{algo2} shows the working methodology of RECOUP routing protocol when a non-LBR/root node ($X$) has a data packet to send to a multicast destination address. When $X$ sends a multicast packet (say $MP_i$) as a source node, it goes through the following steps.

\begin{figure*}[h!]
\centering
  \includegraphics[scale = 0.32]{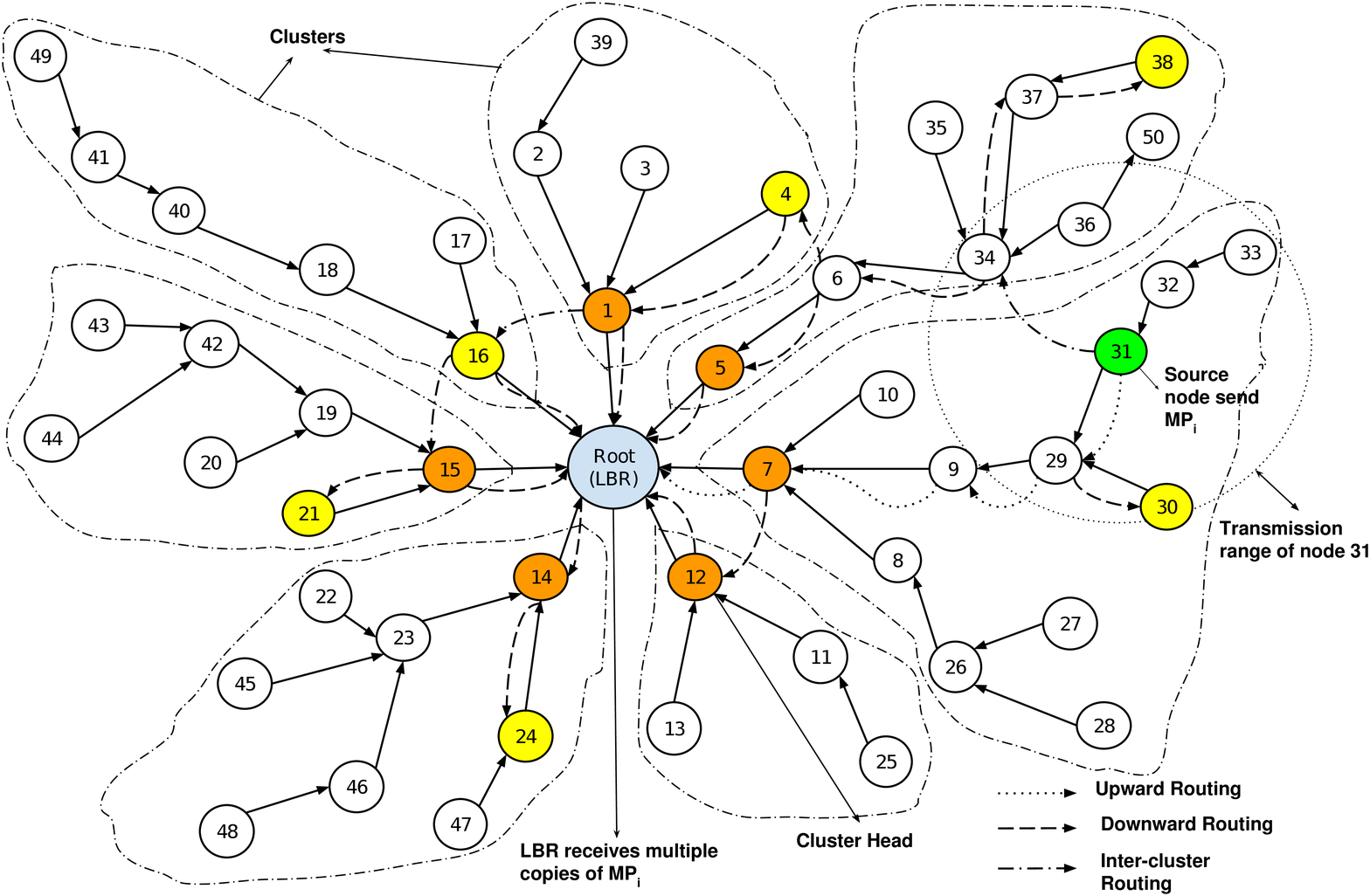}
\caption{\small{Example: RECOUP Cluster Formation and Forwarding Mechanism}}
  \label{Fig:RECOUP}
\end{figure*}

\begin{itemize}
    \item $X$ set the $Pkt_{drop}$ (i.e., maximum number of forwarding hop-counts for $MP_i$ in inter-cluster routing) equal to the rank of $X$ ($X_r$). This is done to avoid the routing loops in the network and to control the number of forwards of $MP_i$ which minimizes the duplicates as well as congestion in the network. The reason that we set $Pkt_{drop}$ to $X_r$ is because after this many number of hops the $MP_i$ will reach to the LBR, and then the LBR could simply send the packet to the remaining interested children using downward multicast routing. The $Pkt_{drop}$ is set by the source node only, and it can be updated by intermediate nodes which decrements it until the value reaches to 0 (refer lines 17 to 19 in Algorithm~\ref{algo2}).   
    
    \item $X$ set the $C_{travel}$ that consists of a set of cluster IDs of the clusters in which $MP_i$ is already forwarded. At the beginning of the routing of $MP_i$ (i.e., when it is at the source node), the $C_{travel}$ at node $X$ contains the ID of $X's$ cluster and the ID of the clusters to which the neighbour nodes of $X$ belongs (refer line 4 in Algorithm~\ref{algo2}). We do not consider the siblings neighbours because all the siblings have the same cluster ID. If a node has no neighbour than its $NC_{ID}$ set remains empty. The intermediate nodes keep on updating the $C_{travel}$ in $MP_i$ with new cluster IDs before forwarding the $MP_i$ to the nodes with cluster IDs that are not present in current $C_{travel}$. Both, the $Pkt_{drop}$ and the $C_{travel}$ values are added in the RPL Packet Information field which is given in IPv6 header format.  
    
    \item Once $X$ sets the $Pkt_{drop}$ and $C_{travel}$, it transmit the $MP_i$ to its preferred parent (P), interested children (IC), and the neighbours ($N_j$) with different cluster IDs. In case there are more than one neighbours of $X$ that belongs to a same cluster, $X$ randomly send the $MP_i$ to only one neighbour from that cluster. This is because the other nodes will receive the $MP_i$ when the intra-cluster\footnote{the intra cluster routing consists of the traditional upward and downward routing techniques that are used for data transmission in RPL's storing mode, i.e., MOP3.} routing is performed for $MP_i$.
\end{itemize}

\par In general, to decrease the number of duplicates in the network, a node (including LBR) never forwards a packet to the node or cluster from which the packet has been received. As it can be seen from Algorithm~\ref{algo2} that if the received packet is not a duplicate, the node perform the following steps.

\begin{itemize}
\item {If the packet is received from a Preferred Parent (refer lines 7 to 12 in Algorithm~\ref{algo2}):}
\begin{itemize}
\item \textit{Step1:} node checks the packet header for $C2C$ flag status, if the flag is not set (i.e., 0), the node is allowed to forward packet to its neighbours that belong to different clusters, else node goes to Step 2.
\item \textit{Step2:} node checks its routing table (or multicast group subscription table) for any interested children that are registered for the multicast address specified in the received packet and then forward packet to them by using OFM, else the node goes to Step 3.
\item \textit{Step3:} if the node itself is a member of the multicast group given in the received packet, then send the packet up to the network stack else discard the packet.
\end{itemize}

\item {If the packet is received from a neighbour/children (refer lines 16 to 12 in Algorithm~\ref{algo2}):}
\begin{itemize}
\item \textit{Step4:} node forward the packet to its preferred parent.
\item \textit{Step5:} node perform the aforementioned steps 1, 2 and 3.
\end{itemize}

\item If a non-root source mote wants to send packet(s), then it will execute the above mentioned steps 1, 2 and 4.  
\end{itemize}

\subsubsection{Example of RECOUP Routing Procedure}
For better understanding routing process of RECOUP, let's consider an IoT network scenario as depicted in Figure~\ref{Fig:RECOUP}. The Figure~\ref{Fig:RECOUP} show the network state after completion of the DODAG formation and cluster creation phase while executing RECOUP protocol in the \textit{Contiki Cooja} emulator. In Figure~\ref{Fig:RECOUP}, nodes 1, 5, 7, 12, 14, 15 and 16 are the cluster heads of the DODAG, and node 31 is the source node of a multicast group which also include nodes 4, 16, 21, 24, 30 and 38. Assume that the cluster IDs of the clusters is same as the node ID of the clusterhead.

\par Node 31 execute the multicast packet transmission procedure as follow: (i) it transmit the multicast packet to its preferred parent i.e., 29; (ii) their are no interested children so it will not send the packet to 32; (iii) in its neighbour set their are only two nodes that does not belong to node 31's cluster, i.e., 34 and 36, but both these nodes belong to the same cluster, so 31 will forward the packet to 34 as it has higher rank than 36. As shown in Figure~\ref{Fig:RECOUP} that from 29 the packet travel towards the root by executing steps 1-4, meanwhile serving all the destination nodes (if any, such as 30) in the way. However, at the same time (i.e., while traveling towards the root) the packet is also disseminated in the various clusters. For instance, upon reception of the packet from node 31, 34 forward it to 6, and it forward the packet to the neighbour cluster by sending it to 4. Node 34 also send the packet towards 38 through 37 because it is registered with the multicast address given in the packet. In this way, the packet travels vertically as well as horizontally at the same time, thus it deceases the propagation delay of the packet for their destinations. 

\par It is seen in Figure~\ref{Fig:RECOUP} that the total number of transmissions ($TX_n$) required to send a multicast packet from node 31 to all its destinations using RECOUP routing mechanism is 13. While, by using the BMRF and RPL MOP~3 the $TX_n$ required is 17. However, the end-to-end propagation delay is not directly proportional because the packet is travelling in various clusters in parallel. For instance, node 29 forwards the packet to node 30 in parallel when 34 forwards it to 37. The value of $TX_n$ greatly depends on the network topology (i.e., DODAG formation) and the position of source and destination nodes. For example, if we add the node 34 in the multicast group and remove 38, then the $TX_n$ required for RECOUP will decrease by a value of two, i.e., $TX_n = 11$, while for BMRF and RPL MOP~3 it will increase by three, i.e., $TX_n = 20$. We are using $TX_n$ parameter because it affect various other metrics that define the communication reliability and scalability in a network. In particular, lower $TX_n$ implies low end-to-end delay and inherent routing support for scalability. 

\subsubsection{Optimized Forwarding Mechanism}

The RECOUP protocol uses ``Optimized Forwarding Mechanism'' (OFM) to minimize forwarding of messages during downward routing. In particular, when a parent receives a multicast packet, and it has $n$ number of interested children for the packet, the parent need to decide whether to send the packet to each children using unicast mode (i.e., create $n$ packets and send one to each child) or to perform a broadcast and all its children will receive the packet. The trade-off between unicast and broadcast mode occurs because the use of unicast mode require more energy consumption as same message is sent by parent for $n$ number of times, while the broadcast makes all the children (including the non-interested ones) to receive the packet in one transmission.

\par The RECOUP protocol uses OFM as follow:  
\begin{itemize}

\item Upward forwarding is done using Link Layer unicast because a node has only one preferred parent at any time during communication process. The inter-cluster forwarding is also done by Link Layer unicast as only one node from a neighbouring cluster needs to receive the message to circulate it in the whole cluster.
\item Downward forwarding is done based on the Mixed mode decision algorithm proposed in BMRF~\cite{Lorente2017} with three as a threshold value in mixed mode.
\end{itemize}

\begin{table}[h!]
\centering
\caption {\footnotesize{Simulation Parameters for RECOUP Protocol Evaluation }}
\scalebox{0.87}{
\begin{tabular}{|p{3cm}|p{6cm}|}
 \hline
\textbf{Parameters} & \textbf{Values} \\ \hline
Emulator & Cooja on Contiki v2.7 \\ \hline
Simulation time & 10 Minutes \\ \hline
Scenario Dimension & 200 x 200 to 800 x 800 sq.meter \\ \hline
Node distribution & Random \\ \hline
Number of nodes & 51 to 201 sky motes (including root) \\ \hline
Transport layer protocol & UDP \\ \hline
Number of source motes & 8 \\ \hline
Routing Protocols & ESMRF, BMRF, and RECOUP\\ \hline
Root waiting timer $t$ & Depends on the value of $\alpha$ \\ \hline
Multicast group or Subscriptions & 20\%, 40\%, 60\%, 80\% \\ \hline
Radio Medium & Unit Disk Graph Medium (UDGM) \\ \hline
PHY and MAC Layer & IEEE 802.15.4 with CSMA and ContikiMAC \\ \hline
RNG Seed & 30 iterations each with new seed \\ \hline
Application protocol & CBR \\ \hline
Transmission Range  & 25m \\ \hline
Number of attacker nodes & 10\% to 40\% \\ \hline
Traffic rate & 0.50 pkt/sec - 500 packets \\ \hline
ESMRF & Contiki v2.7 Default Configuration \\ \hline
BMRF and RECOUP & Mixed mode (Threshold: 3) \\ \hline
\end{tabular}}
\label{tab:simu} 
\end{table}

\section{Simulation and Result Evaluation}
\label{sec:result}
In this section, we present the performance evaluation of our proposed RECOUP protocol by using the simulation results. We have compared the performance of RECOUP protocol with the state-of-the-art protocols that includes ESMRF~\cite{Fadeel2015}, a enhanced stateless multicast RPL-based forwarding protocol, and BMRF~\cite{Lorente2017}, a bidirectional multicast RPL forwarding protocol. We have fully implemented RECOUP protocol on \textit{Cooja}, the Contiki network emulator~\cite{Dunkels}, and we used the available open source codes of ESMRF and BMRF for comparison purposes. Table~\ref{tab:simu} provide the details of various parameters along with their values that we have used to configure the target IoT network scenarios in \textit{Contiki Cooja} emulator.   
\subsection{Performance analysis}
In this section, we present the performance analysis of our proposed protocol in terms of various metrics and compare it with the most relevant and recent state-of-the-art works. The evaluation metrics that we have used are as follow: (i) packet delivery ratio, (ii) end-to-end delay, (iii) per packet energy consumption, and (iv) memory consumption. The values for these metrics are calculated in different IoT scenarios that are created by varying network size, network traffic, and attacker nodes. The attacker nodes could perform either the rank attack or the blackhole attack to disrupt data communication process. The source and multicast destination nodes are selected randomly, and the final results plotted are the average of 30 simulation runs each with different seed value.

\begin{figure}[ht!]
\centering
  \includegraphics[scale = 0.64]{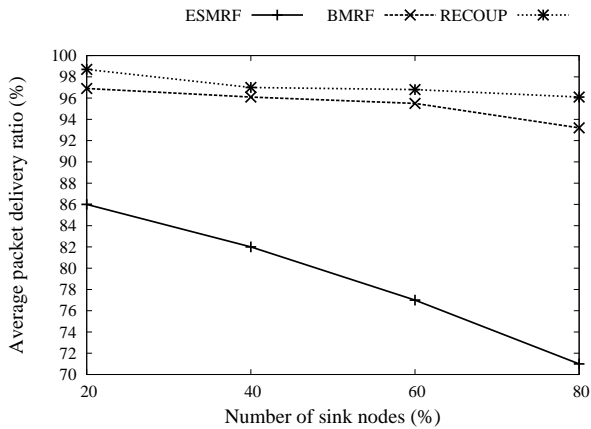}
\caption{\small{Packet delivery ratio with increased percentage of sinks}}
  \label{fig:loadpdr1}
\end{figure}

\begin{figure}[ht!]
\centering
  \includegraphics[scale = 0.64]{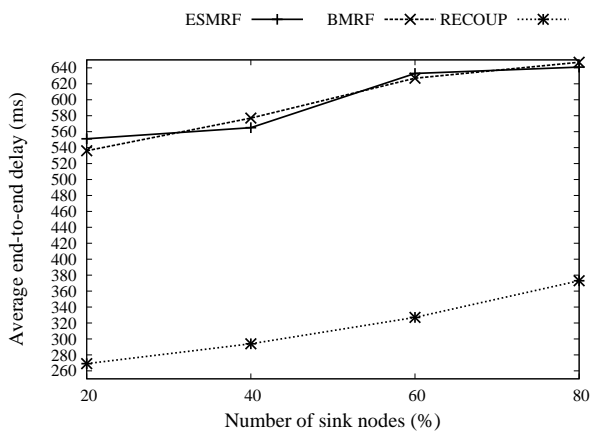}
\caption{\small{End-to-end delay with increased percentage of sinks}}
  \label{fig:loaddelay1}
\end{figure}

\begin{figure}[ht!]
\centering
  \includegraphics[scale = 0.64]{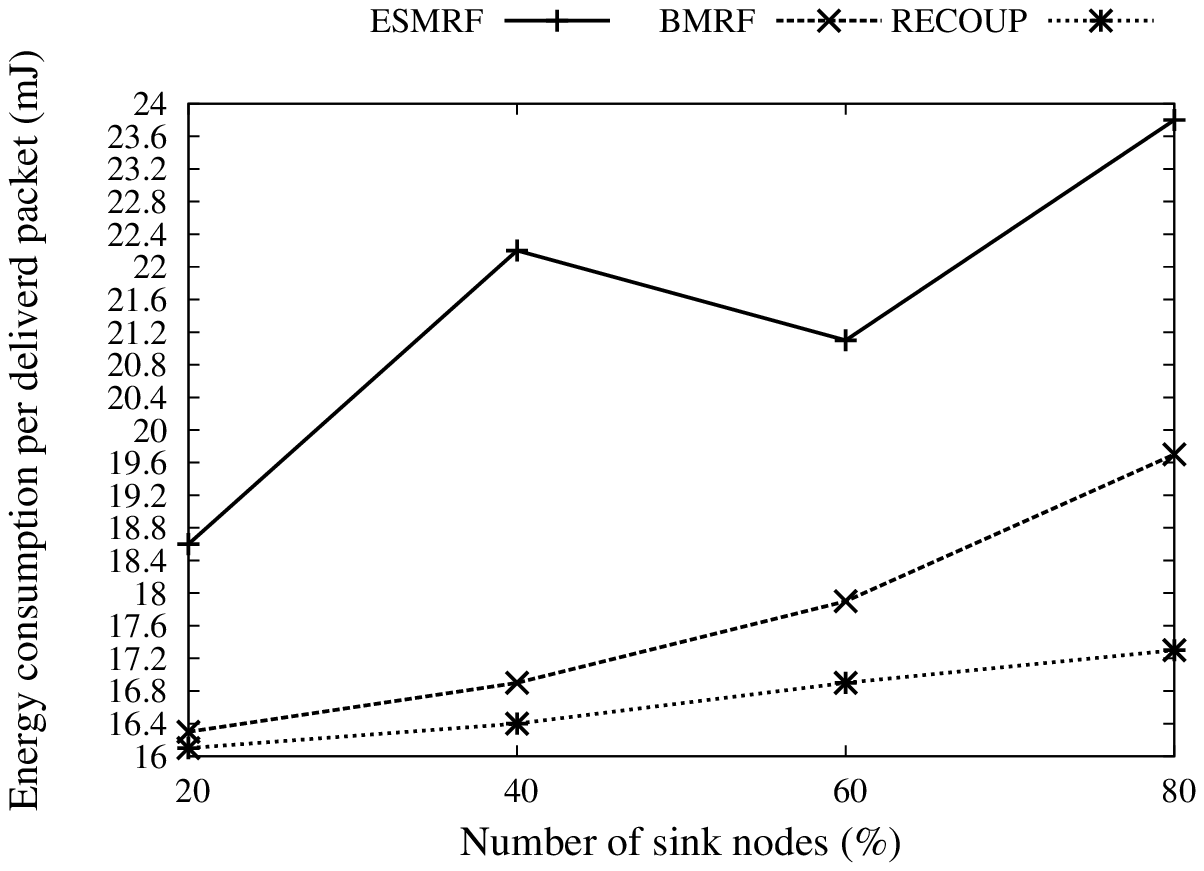}
\caption{\small{Energy consumption with increased percentage of sinks}}
  \label{fig:loadenergy1}
\end{figure}

\begin{figure}[ht!]
\centering
  \includegraphics[scale = 0.64]{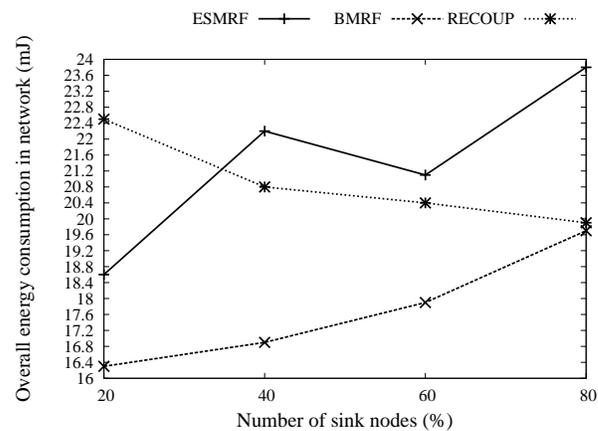}
\caption{\small{Overall energy consumption with increased percentage of sinks}}
  \label{loadenryall1}
\end{figure}

\subsubsection{Effect of increase in network load}

In this section, we discuss how the various networking parameters (e.g., packet delivery ratio, end-to-end delay, and energy consumption) are influenced by the change in the percentage of subscribers (or sinks/destinations) in the target network for the ESMRF, BMRF and RECOUP routing protocols. It should be noted that in the target network scenario we are varying the number of sink node's percentage in the range of 20\% to 80\% but keeping the fixed number of source nodes (i.e., 8). With the increase in the sink nodes in the network, the network traffic increases as more number of packets are travelling (depending upon the location of the destination) in the network. For these scenarios we assume that no adversary is present in the network, and the network consists of 100 nodes (excluding the LBR/root). 

\par Figure~\ref{fig:loadpdr1} shows the change in the average Packet Delivery Ratio (PDR) for all the comparing protocols with increase in the percentage of sink nodes. As shown in Figure~\ref{fig:loadpdr1}, the PDR of RECOUP remains higher as compared to ESMRF and BMRF. However, BMRF and RECOUP has more or less have same the PDR due to their upward and downward forwarding mechanism, RECOUP has slightly higher PDR then BMRF due to its inter-cluster forwarding rule. The inter-cluster forwarding helps RECOUP to disseminate the packets even in (small) partitioned network areas, additionally it forward the packets by going around the broken or weak links which might have been created due to node transmission range or interference. ESMRF has the lowest PDR due to its strict upward and downward forwarding mechanism which increases the number of transmission required to send the packet to all the destinations.

\par Figure~\ref{fig:loaddelay1} depicts the effect on average End-to-End Delay (EED) for all the comparing protocols with increase in the percentage of sink nodes. It can be seen from the figure that RECOUP remains too low when compared with the ESMRF and BMRF. This is because in RECOUP for most of the times the packets reaches to its intended destinations without travelling through root. For instance in Figure~\ref{Fig:RECOUP}, the node 38 receive all its packets in three transmissions while for ESMRF and BMRF the packets will have to travel 9 hops before it reaches to node 38. In particular, the use of efficient inter-cluster forwarding along with the upward and downward forwarding of multicast messages in RECOUP protocol triggers a quick dissemination of packets in the whole network in no time. Due the aforementioned reason the energy consumption for per packet (ECP) delivery is also lower in RECOUP as it is shown in Figure~\ref{fig:loadenergy1}. In Figure~\ref{fig:loadenergy1}, BMRF has lower energy consumption then ESMRF due to two reasons, first it uses the optimized forwarding scheme and second it serves the destinations while forwarding a packet towards root node.     

\par Although, the ECP of RECOUP is lower than other comparing protocols, but same is not true when the energy consumption of the whole network is calculated, this is depicted in Figure~\ref{loadenryall1}. This is because the number of packet transmissions are higher in RECOUP. This is because in RECOUP a packet might have to travel in a neighbour cluster even in the cases in which no multicast member(s) resides in that cluster. This is the cost that RECOUP has to pay to achieve improved network reliability and resistance to an array of routing attacks. However, as the network load increases the additional energy consumption with respect to the network throughput will start decreasing as more and more sinks will benefited by RECOUP's inter-cluster routing scheme.

\begin{figure}[ht!]
\centering
  \includegraphics[scale = 0.64]{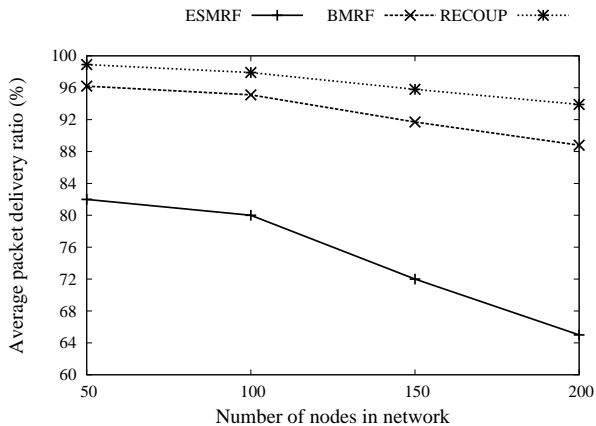}
\caption{\small{Packet delivery ratio with increased network size}}
  \label{fig:loadpdr}
\end{figure}

\begin{figure}[ht!]
\centering
  \includegraphics[scale = 0.64]{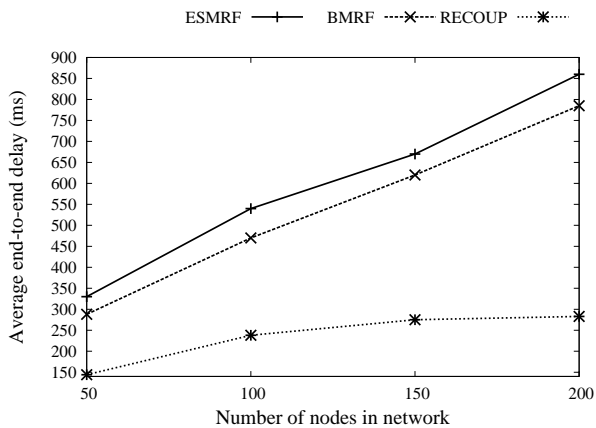}
\caption{\small{End-to-end delay with increased network size}}
  \label{fig:loaddelay}
\end{figure}

\begin{figure}[ht!]
\centering
  \includegraphics[scale = 0.64]{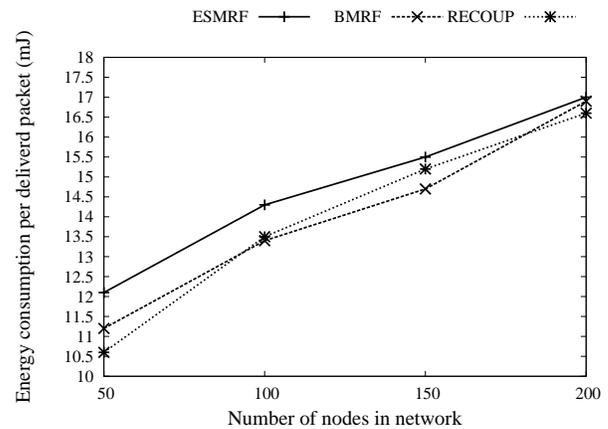}
\caption{\small{Energy consumption with increased network size}}
  \label{fig:loadenergy}
\end{figure}

\subsubsection{Effect of increased in network size}

In this section, we present the effect of increase in the network size for all the comparing protocols. This is important for the applications where scalabilty of the network is an important factor. In this scenario, we fix the number of source (i.e., 8) and sinks (i.e., 40\%).

\par The effect on PDR with increase in network size for all the comparing protocols in shown in Figure~\ref{fig:loadpdr}. It can be seen in Figure~\ref{fig:loadpdr} that BMRF and RECOUP protocols remain less affected by the increase in network size when compared to ESMRF. With the increase in network size, the number of transmissions and number of hops between the source and destinations increase greatly, however, due to inter-cluster routing the increase in number of intermediate hops in RECOUP is not to high, which helps it to keep the PDR higher even in large networks. The results for the EED in Figure~\ref{fig:loaddelay} also support the aforementioned reason as it shows the lower increase in EED for RECOUP with increased network size. 

\par The change in the energy consumption for per packet (ECP) delivery for the comparing protocols with increase in network size is shown in Figure~\ref{fig:loadenergy}. As the network size increases, the number of duplicate packets also increases for the RECOUP protocol, thus, the ECP increases. It is because we are also taking into account the energy consumed by receiving of a duplicate packet at a destination node. The increase in ECP for ESMRF and BMRF is mainly due to increase in the number of transmissions required for a packet to reach to its destinations.

\begin{figure}[ht!]
\centering
  \includegraphics[scale = 0.64]{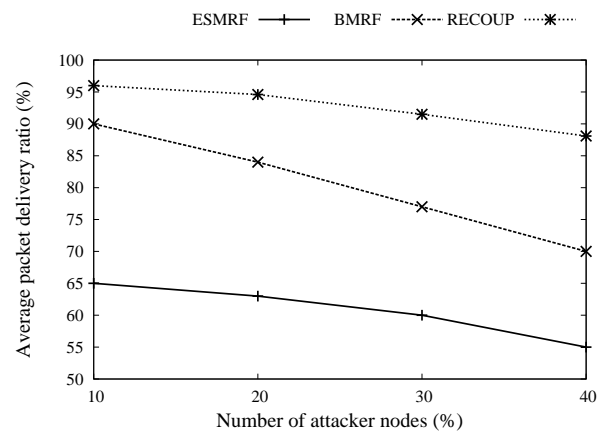}
\caption{\small{Packet delivery ratio with increased percentage of attackers}}
  \label{fig:attpdr}
\end{figure}

\begin{figure}[ht!]
\centering
  \includegraphics[scale = 0.64]{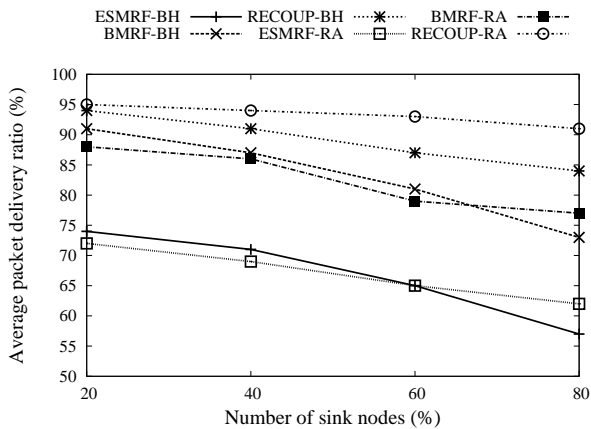}
\caption{\small{Packet delivery ratio of comparing protocols in presence of rank and blackhole attacks}}
  \label{fig:attpdrcomp}
\end{figure}


\begin{figure}[ht!]
\centering
  \includegraphics[scale = 0.64]{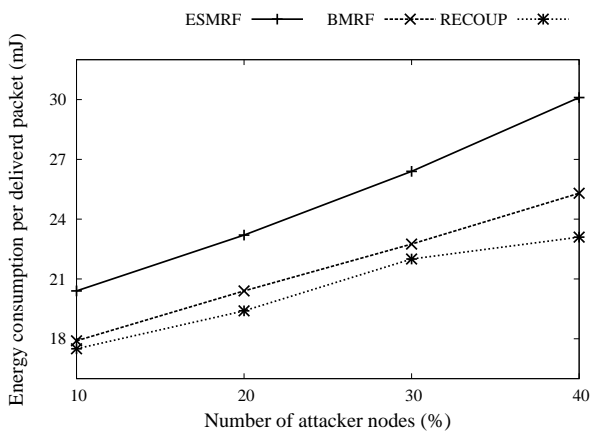}
\caption{\small{Energy consumption with increased percentage of attackers}}
  \label{fig:attenergy}
\end{figure}

\subsubsection{Effect of increase in number of attacker nodes}
In this section, we evaluate the performance of RECOUP in presence of attacker nodes. We randomly configure nodes to either perform the rank attack followed by selective packet discarding or the blackhole attack. The rank attack will disrupt the correct formation of DODAG, thus it creates weak or broken links in the network, while the blackhole attacker will drop all the received packets without forwarding, thus it decreases the packet delivery ratio. Figure~\ref{fig:attpdr} show the effect on PDR with increase in number of attackers for all the comparing protocols. In this scenario, we set the network size to 101 nodes with 8 source nodes and 40\% sink nodes. It can be seen from Figure~\ref{fig:attpdr} that RECOUP has minimal effect of the attacker presence due to its inter-cluster forwarding mechanism. For instance, while using ESMRF and BMRF protocols, if node 29 behaves like a blackhole attacker in the topology given in Figure~\ref{Fig:RECOUP}, then all the packets sent by source node 31 will never reach to its destination nodes. Alternatively, if RECOUP is running as the routing protocol, then all its destinations (except node 30) will receive the packets sent by 31. This behaviour of RECOUP greatly increase its PDR even in the presence of attackers. Although RECOUP only provides communication robustness in presence of attackers, and it does not detect the attackers, but it can be easily done by using a traffic analysis tool at LBR, which when see that the packets sent by 31 are received through other clusters but not from its own cluster could generate a security alarm. Due to the aforementioned functionality of RECOUP, we can see in Figure~\ref{fig:attpdrcomp} that RECOUP has the highest PDR for both attacks when it is compared with ESMRF and BMRF protocols. It has been also seen that the position of the attacker greatly affects the PDR as the attacker near the root is much more effective when compared with the attackers residing close to leaf nodes.   

\par Although we have only tested RECOUP in presence of rank and blackhole attacks, but from the functionality of RECOUP it is clear that it can effectively minimize the effects of other routing attacks such as jamming and sybil attacks. It is because in RECOUP the packet travels through multiple routes toward its destination, hence the failure or maliciousness of few nodes or links won't effect much to its routing process.

\par Figure~\ref{fig:attenergy} shows how the increase in number of attackers affect the ECP delivery for all the three comparing protocols. The ECP for ESMRF, BMRF, and RECOUP protocols increases with the increase in the attackers in the network. It is because the presence of attackers (mainly the rank attacker) on routing paths increases the number of hop-to-hop re-transmissions. Additionally, the increase in routing path length and disruption in DODAG creation caused by rank attack will leave the network with non-optimal routes. As the RECOUP does not provide any mitigation to these attacks, the increase in the energy consumption per packet shows the same trend for all the protocols.

\begin{table}[h!]
\centering
\caption {\footnotesize{Memory Usage}}
\scalebox{1.05}{
\begin{tabular}{|c|c|c|}
 \hline
\textbf{} & \textbf{Flash [Bytes]} & \textbf{RAM [Bytes]} \\ \hline
ContikiRPL & 41498 & 8246  \\ 
RECOUP & 488 (+1.2\%) & 292 (+3.5\%)\\ \hline
Total & 42170 (+1.2\%) & 8574 (+3.5\%)\\ \hline
\end{tabular}}
\label{tab:memory} 
\end{table}

\par Finally, Table~\ref{tab:memory} shows ContikiRPL memory consumption~\cite{SKY} and overall code and data memory increase when implementing RECOUP. The memory consumption in RECOUP is slightly higher than the state-of-the-art protocols. In order to correctly implement all the functionalities of RECOUP protocol, following additional information is stored at a node: (i) to perform the duplication detection, a node needs to create and maintain, the $DD_{tab}$, this table consists of three fields (i.e., message sequence number and source-destination address pairs); and (ii) a new field is added in neighbour table which consists of entries of its neighbours Cluster ID ($C_{id}$). The cost of RECOUP is $488$ Byte of Flash and $292$ Byte of RAM. We consider around 95 to 100 entries for both the tables, which are sufficiently large amount w.r.t a large IoT network. The additional memory consumption in RECOUP compared to the traditional RPL protocol is almost negligible considering the additional features it provides.

\subsubsection{Discussion on data communication reliability and security} 

The optimized inter-cluster forwarding mechanism used in RECOUP greatly reduces single point of failures in the network, thus it makes the communication system more robust and reliable for data communications. All the existing multicast routing approaches suffer from the scalability issue. It is because as the number of nodes increases in the network, the size of DODAG tree increases which causes increase in the number of hops traveled by a message and it decreases the packet delivery ratio. It further increases the following: (i) probability of a route break, (ii) the energy consumption, and (iii) the end-to-end delay. Apart from the end-to-end delay, the PDR is a critical metric in various application scenarios where sensitive operations are dependent on the information received from other parts of the network. Hence, we believe that providing communication reliability along with the network scalability while keeping in mind the constrained nature of IoT devices is a major challenge for routing protocols in IoT networks.         

\par Communication security is considered as one of the key challenges in IoT networks due to its openness. In~\cite{Wallgren2013}, authors discuss few of the well known security attacks on the RPL protocol which includes selective-forwarding, sinkhole, blackhole, wormhole, clone ID or sybil, and rank attacks~\cite{Glissa2016}. Their research shows that the RPL protocol running on top of 6LoWPAN networks is vulnerable to all the aforementioned attacks. Furthermore, all the existing extensions of RPL which includes the ESMRF and BMRF also fails to address any of these security threats in IoT networks. It is because all these protocols use the DODAG topology which contains the possibility of single point of failure. For instance, in Figure~\ref{Fig:RECOUP}, if node 9 is down due to some technical fault or an attack performed by some adversary, all the messages sent by it will never reach to its destination nodes (i.e., 4, 16, 21 and 24).

\par In IoT network scenarios, depending upon the application requirements, we might have real time deadlines. However, the devices are deployed in an insecure environment, thus ensuring the communication reliability, and on-time and secure communication are crucial aspects. To this end, RECOUP performs the data communication in a way that ensures that it will avoid the single point failures, push the network communication towards scalability, minimize the effects network partitioning, and reduce the propagation delay for recipients of the data packets. Rather than getting failed in an IoT environment, our protocol works reasonably better and send its data traffic successfully. Due to the inter-cluster communication which triggers a faster dissemination of the information, RECOUP is able to easily mitigate the worst effects of few of the aforementioned attacks. The scalability and quick dissemination features of RECOUP could also be very helpful in enhancing the performance of large scale attestation techniques~\cite{pals2018} used in IoT networks. These attestation techniques dynamically verifies the integrity of various software and hardware components residing on an IoT device at runtime which increases the network security.

\section{Conclusions}
\label{sec:conclusions}

In this paper, we propose a novel reliable and robust multicast routing protocol (RECOUP) for Low-power and Lossy Networks such as 6LoWPAN, which are highly used for deploying IoT networks for various smart service applications. RECOUP uses the advantages of the recently proposed BMRF protocol which includes support for dynamic group registrations and enabling upward and downward forwarding, and it addresses BMRF's key disadvantages such as higher end-to-end delay, low security against single point of failures, and low scalability. We show that RECOUP remains largely unaffected from rank and blackhole attacks as it delivers more than 80\% of data packets to destinations in the presence of attackers. From the simulation results, we can conclude that RECOUP effectively achieve its goals (reliability and robustness) at the expense of a slightly higher energy and memory consumption. Please note that RECOUP does not aims to provide explicit security solutions for various attacks (e.g., blackhole, rank attack, selective packet discarding, wormhole, etc), however, it proves that it is robust and effective in data communication process in the presence of these attacks.  
\par As the RECOUP protocol only provide resistance for the security attacks but not the mitigation, in the future work, from the security point of view we are looking to embed algorithms to countermeasure more specific (to IoT network) attacks such as rank attacks.    

\section*{Acknowledgements}
Pallavi Kaliyar is pursuing her Ph.D. with a fellowship for international students funded by Fondazione Cassa di Risparmio di Padova eRovigo (CARIPARO). This work is also supported in part by EU LOCARD Project under Grant H2020-SU-SEC-2018-832735. The work of M.  Conti was supported by the Marie Curie Fellowship through European Commission under Agreement PCIG11-GA-2012-321980.

\bibliographystyle{IEEEtran}
\bibliography{uw_security}

\end{document}